\documentclass[aip,amsmath,amssymb,reprint,floatfix]{revtex4-2}
\usepackage{url}

\usepackage{microtype}
\usepackage{graphicx}
\usepackage{dcolumn}
\usepackage{bm}

\usepackage[T1]{fontenc}
\usepackage{mathptmx}
\usepackage{etoolbox}

\makeatletter
\def\@email#1#2{%
 \endgroup
 \patchcmd{\titleblock@produce}
  {\frontmatter@RRAPformat}
  {\frontmatter@RRAPformat{\produce@RRAP{*#1\href{mailto:#2}{#2}}}\frontmatter@RRAPformat}
  {}{}
}%
\makeatother
\begin{document}

\preprint{AIP/123-QED}

\title{Role of Excited States in Resonant Charge Transfer during Li$^+$ Backscattering from MoS$_2$: A Multi-Orbital Theoretical Study}
\author{T. A. Balsamo}
\affiliation{Dpto. de Física, Facultad de Ingeniería Química, Santiago del Estero 2829, Santa Fe, Argentina.}
\author{F. G. Ibarlucea}
\affiliation{Dpto. de Física, Facultad de Ingeniería Química, Santiago del Estero 2829, Santa Fe, Argentina.}

\author{M. A. Romero}
\email{marcelo.romero@santafe-conicet.gov.ar}
\affiliation{Dpto. de Física, Facultad de Ingeniería Química, Santiago del Estero 2829, Santa Fe, Argentina.}
\affiliation{Instituto de Física del Litoral (UNL-CONICET), Güemes 3450, Santa Fe, Argentina.}

\date{\today}

\begin{abstract}
We present a theoretical investigation of resonant charge transfer in low-energy Li$^+$ ions backscattered from a MoS$_2$ surface, focusing on the influence of excited projectile states. Using a time-dependent Anderson model in the infinite-$U$ limit, we evaluate the individual contributions from the Li 2\textit{s}, 2\textit{p$_x$}, 2\textit{p$_y$}, and 2\textit{p$_z$} orbitals to the final charge state distribution. The Hamiltonian parameters are computed using an expanded Huzinaga basis set that explicitly incorporates lithium's 2\textit{p} orbitals. Each orbital channel is treated separately, and electronic correlation effects are introduced approximately through a probabilistic exclusion principle applied to the final charge state. Theoretical calculations demonstrate that including 2\textit{p} channels enhances the agreement with previously reported experimental neutral fractions, where the single-channel descriptions show noticeable discrepancies. Among excited states, the 2\textit{p$_z$} orbital oriented perpendicular to the surface contributes most significantly due to its spatial extension toward the substrate. Examination of temporal evolution reveals that independent channel occupations exceed unity at small projectile$-$surface separations, highlighting the necessity of dynamical correlation treatments. These findings establish that excited states make non-negligible contributions to charge exchange processes for alkali ions interacting with transition metal dichalcogenide surfaces and should be incorporated for accurate quantitative modeling.
 
\end{abstract}

\maketitle

\section{\label{sec:intro}Introduction}

Layered two dimensional materials, particularly transition metal dichalcogenides (TMDs) such as molybdenum disulfide (MoS$_2$), have become cornerstones of modern nanoscience and nanotechnology. In its bulk 2H semiconducting phase, MoS$_2$ exhibits chemical stability, a van der Waals surface, and unique electronic characteristics, making it an ideal platform for studying surface mediated charge transfer and for applications in optoelectronics, catalysis, and energy storage \cite{adma2018,CHEN2022,ma1612,XIE2026140406}. 

Among experimental techniques, low energy ion scattering (LEIS) spectroscopy has established itself as a valuable tool for examining charge transfer, owing to its inherent surface sensitivity and its ability to resolve the charge states of scattered particles \cite{BRONGERSMA,C6AY00765A,PRUSA2024}. Alkali projectiles such as Li$^+$ introduce additional richness because their low ionization potential places the active level in close resonance with the valence band of many semiconductors, favoring resonant charge exchange \cite{PhysRevA032901,BORISOV2000430,PhysRevLett196102,Canario2006,PhysRevB075422,GARCIA2009597,PhysRevA052901,PhysRevB165307,jp4116673,PhysRevB125411}. This allows experimental observation of positive, neutral, and even negative outgoing species.

In a recent combined experimental and theoretical investigations \cite{1010630283586}, the authors examined Li$^+$ collisions with MoS$_2$ surfaces across the 2.5--8.0 keV energy range, revealing neutralization fractions between 20 \% and 35 \% for projectiles scattered from Mo atoms. Theoretical modelling using the infinite-$U$ Anderson Hamiltonian with a restricted lithium basis set (incorporating only the 2\textit{s} orbital) successfully reproduced the magnitude of the experimental observations, but not their trend, and systematically underestimated neutralization probabilities at higher incident energies (above 6.0 keV). This discrepancy was tentatively attributed to the omission of excited projectile states (particularly the Li 2\textit{p} manifold) which may become accessible under conditions of stronger coupling and shorter interaction times \cite{WINTER2002387}. Indeed, as noted in Ref.~[\onlinecite{1010630283586}], the Li 2\textit{p} level also resonates with the MoS$_2$ valence band and could contribute to electron capture, especially during the final stages of the exit trajectory where the projectile-surface distance is small and the ionization level is significantly broadened.

The present work addresses this limitation by presenting a purely theoretical analysis of charge exchange dynamics in low energy Li$^+$ backscattering from MoS$_2$, concentrating on the role of excited states. We extend previous theoretical treatments by implementing an expanded Huzinaga basis set \cite{huzinaga1984gaussian,Huzinaga2p} for lithium that explicitly incorporates the 2\textit{s}, 2\textit{p$_x$}, 2\textit{p$_y$}, and 2\textit{p$_z$} orbitals as independent charge accepting channels. Charge transfer dynamics are computed for each channel separately within the infinite-$U$ Anderson framework. Electronic correlation is incorporated approximately through a probabilistic exclusion scheme applied to final charge states, which respects the constraint that at most one electron (or hole) can occupy the projectile. This methodology allows us to quantify each orbital channel's contribution to the overall neutralization and to assess the importance of excited states in the complete charge exchange process.

Our results demonstrate that incorporating the 2\textit{p} channels significantly improves agreement with experimental data, particularly at higher energies where the single channel (2\textit{s} only) calculations deviate. Among the excited orbitals, the perpendicularly oriented 2\textit{p$_z$} component displays the strongest contribution, owing to its favourable spatial extension toward the substrate, which enhances wavefunction overlap with surface states. Temporal evolution analysis reveals that  dynamical correlation effects are necessary for a fully accurate description.

These findings underscore the importance of including excited states in theoretical models of charge exchange for alkali ions on semiconductor surfaces. They also provide a more complete understanding of the microscopic mechanisms governing ion surface interactions on two dimensional materials such as MoS$_2$, and offer a pathway to refine LEIS based investigations of surface electronic structure.

The remainder of this paper is organized as follows. Section  \ref{sec:2} presents the theoretical framework, including the Anderson Hamiltonian, the Green function formalism, the ab initio determination of Hamiltonian parameters, the trajectory model, and the multi-channel correlation approximation. Section \ref{sec:3} presents our results, including the effects of basis set choice, the energy levels and hybridization widths, the independent and correlated neutralization channels, negative ion formation, and the temporal evolution along trajectories. Section \ref{sec:4} summarizes our conclusions and outlines directions for future work.

\section{THEORETICAL FRAMEWORK FOR RESONANT CHARGE TRANSFER} \label{sec:2}
The charge exchange process occurring during the scattering of low-energy ions from a solid surface is described within the framework of resonant charge transfer (RCT). In this mechanism, the energetic position of the projectile's electronic level shifts dynamically as it approaches and recedes from the surface due to its interaction with the continuum of electronic states characteristic of the target. Depending on the overlap between the ion's active level and the local density of states (DOS) of the surface (as well as the time-dependent coupling between them) electrons may tunnel into or out of the projectile, thereby altering its charge state throughout the interaction.

To model this process, we employ a quantum mechanical approach based on the Anderson-Newns Hamiltonian \cite{PhysRev.124.41}, widely applied in studies of charge exchange between atoms and surfaces, in both equilibrium \cite{PhysRevB74195419,Romero2012,GARCIA2018507,PhysRevB83125411} and non-equilibrium \cite{PhysRevA.103.062805,doi:10.1021/acs.jpcc.9b10042,LUNA2008237,VIDAL201118,PhysRevA032901,PhysRevA.89.042702,doi:10.1021/jp511339v,PhysRevB.93.195439,ROMERO2022122070,PhysRevA110012806,ROMERO2025163604,PhysRevB.96.075424,Romero2009,PhysRevB87195419,doi:10.1021/acs.jpcc.8b09828,PhysRevB075422,GARCIA2009597,jp4116673,Bonetto_2007,https://doi.org/10.1002/pssa.200304907,PhysRevB.72.035432,GARCIA20062195} processes. This theoretical framework represents the projectile as a localized electronic state coupled to a continuum of surface states. The interaction is treated assuming a head-on binary collision between Li$^+$ and a surface molybdenum atom, consistent with experimental geometries in which only projectiles scattered along the surface-normal direction are detected. Prior investigations indicate that final charge states are predominantly determined during the outgoing trajectory following the point of closest approach \cite{PhysRevA.103.062805,doi:10.1021/acs.jpcc.9b10042,LUNA2008237,VIDAL201118,PhysRevA032901,PhysRevA.89.042702,doi:10.1021/jp511339v,PhysRevB.93.195439,PhysRevB.96.075424,doi:10.1021/acs.jpcc.8b09828}.

In this work, we adopt the simplified infinite-$U$ limit of the Anderson model. While the origin of this approximation is to exclude the doubly occupied configuration, it can also be regarded as an independent channel approximation that incorporates spin fluctuation. Thus, as will be shown below, we can evaluate the negative ion fraction within the same approximation, but in the hole representation. This approximation has been successfully applied to similar systems\cite{PhysRevA032901,PhysRevB.96.075424,PhysRevB075422,jp4116673}, particularly where coupling to the negative ion channel is weak or negligible. Moreover, it allows for a fully time-dependent treatment of the charge exchange dynamics. In contrast to prior treatments restricted to the Li 2s orbital, the present analysis incorporates excited 2\textit{p$_x$}, 2\textit{p$_y$}, and 2\textit{p$_z$} states as independent channels, enabling evaluation of excited-state contributions to the total neutralization probability.

In the following, atomic units (a.u.) are employed unless stated otherwise.
\subsection{Hamiltonian formulation and configuration space}
The Anderson  Hamiltonian \cite{PhysRev.124.41} describes interactions between the Li projectile and MoS$_2$ surface through three distinct components:
\begin{equation}
\hat{H}={{\hat{H}}_{surf}}+{{\hat{H}}_{proj}}(t)+{{\hat{H}}_{coupl}}(t)
\end{equation}
The surface term ${\hat{H}}_{surf}$ represents target energy and is diagonalized independently of projectile-surface interactions:
\begin{equation} \label{eq:2}
{{\hat{H}}_{surf}}=\sum\limits_{\vec{k},\sigma }{{\varepsilon }_{\vec{k}\sigma }\hat{c}_{\vec{k}\sigma }^{\dagger }{{{\hat{c}}}_{\vec{k}\sigma }}}	
\end{equation}	
Here, ${\varepsilon }_{\vec{k}\sigma }$ denotes eigenenergies of the surface electronic states, while $\hat{c}_{\vec{k}\sigma }^{\dagger }$ and ${{\hat{c}}_{\vec{k}\sigma }}$ represent creation and annihilation operators for electrons with spin projection $\sigma$ in state $\vec{k}$ within second quantization.

The remaining terms ${{\hat{H}}_{proj}}$ and ${{\hat{H}}_{coupl}}$  describe projectile energy and its coupling to the surface respectively. Resonant charge transfer to both ground and excited states is considered. The model restricts projectile charge fluctuations to 0 or 1 electron for each independently considered orbital channel $\alpha $ (where $\alpha =2s$, $2{{p}_{x}}$, $2{{p}_{y}}$, $2{{p}_{z}}$). The configuration space for each channel comprises:
\begin{itemize}
  \item Neutral Li$^0$ state with singly occupied orbital $\alpha$ and spin $\sigma =\uparrow ,\downarrow $, denoted $\left| {{\sigma }_{\alpha }} \right\rangle $
  \item Positive Li$^+$ states with empty orbital $\alpha$, denoted as $\left| {{0}_{\alpha }} \right\rangle $
\end{itemize}
Within this restricted independent-channel framework, the Hamiltonian terms take the following forms using projection-operator language:	
\begin{equation} \label{eq:3}
\hat{H}_{proj}^{\alpha }=E_{0}^{\alpha }\left| {{0}_{\alpha }} \right\rangle \left\langle  {{0}_{\alpha }} \right|+E_{1}^{\alpha }\sum\limits_{\sigma }{\left| {{\sigma }_{\alpha }} \right\rangle \left\langle  {{\sigma }_{\alpha }} \right|}
\end{equation}
\begin{equation}\label{eq:4}
\hat{H}_{coupl}^{\alpha }=\sum\limits_{\vec{k},\sigma }{\left[ V_{\vec{k}\alpha }^{\sigma }\hat{c}_{\vec{k}\sigma }^{\dagger }\left| {{0}_{\alpha }} \right\rangle \left\langle  {{\sigma }_{\alpha }} \right|+V_{\vec{k}\alpha }^{\sigma *}\left| {{\sigma }_{\alpha }} \right\rangle \left\langle  {{0}_{\alpha }} \right|{{{\hat{c}}}_{\vec{k}\sigma }} \right]}
\end{equation}

In Eq.~\ref{eq:3}, $E_{0}^{\alpha }$ and $E_{1}^{\alpha}$ represent total energies of Li projectiles with 0 and 1 electron in orbital $\alpha$ respectively, related to orbital ionization energy ${{\varepsilon }_{\alpha }}$ through ${{\varepsilon }_{\alpha }}=E_{1}^{\alpha }-E_{0}^{\alpha }$. Equation~\ref{eq:4} introduces $V_{\vec{k}\alpha }^{\sigma }\left( t \right)$ as the time-dependent coupling between the projectile's active $\alpha$-state and surface state $\vec{k}$.

Each independent channel satisfies the normalization condition:
\begin{equation}
\left| {{0}_{\alpha }} \right\rangle \left\langle  {{0}_{\alpha }} \right|+\sum\limits_{\sigma }{\left| {{\sigma }_{\alpha }} \right\rangle \left\langle  {{\sigma }_{\alpha }} \right|}={{\hat{1}}_{\alpha }}
\end{equation}

The independent-channel approximation treats each orbital $\alpha $ as if it were the only active level on the projectile. This means that for each channel, the projectile can be either empty (positive ion) or singly occupied (neutral atom), but the occupation of different channels is not coupled during the dynamics. This approximation is justified when the coupling between channels is weak, which is the case for the Li $2s$ and $2p$ orbitals due to their different symmetries and energy positions. However, as we will show, the independent-channel results can violate the normalization condition at short distances, necessitating full correlated treatment.

\subsection{Green function formalism for charge fractions}
Time evolution of Li projectile charge states during interaction is described using non-equilibrium Green-Keldysh formalism \cite{Keldysh:1964ud}. The neutralization probability for channel $\alpha $ at time $t$ is expressed as:
\begin{equation}
P_{\alpha }^{0}\left( t \right)=\sum\limits_{\sigma }{{{\left\langle \left| {{\sigma }_{\alpha }} \right\rangle \left\langle  {{\sigma }_{\alpha }} \right| \right\rangle }_{t}}}
\end{equation}
Temporal evolution of ${{\left\langle \left| {{\sigma }_{\alpha }} \right\rangle \left\langle  {{\sigma }_{\alpha }} \right| \right\rangle }_{t}}$ is derived from equal-time Green-Keldysh functions. The required Green function takes the form: 
\begin{equation}\label{eq:7}
F_{\sigma }^{\alpha }\left( t,t' \right)=i\left\langle \left[ \left| {{\sigma }_{\alpha }} \right\rangle {{\left\langle  {{0}_{\alpha }} \right|}_{t'}},\left| {{0}_{\alpha }} \right\rangle {{\left\langle  {{\sigma }_{\alpha }} \right|}_{t}} \right] \right\rangle
\end{equation}
which requires knowledge of the advanced Green function:
\begin{equation}\label{eq:8}
G_{\sigma }^{\alpha }\left( t,t' \right)=i\,\Theta \left( {t}'-t \right)\left\langle \left\{ \left| {{\sigma }_{\alpha }} \right\rangle {{\left\langle  {{0}_{\alpha }} \right|}_{t'}},\left| {{0}_{\alpha }} \right\rangle {{\left\langle  {{\sigma }_{\alpha }} \right|}_{t}} \right\} \right\rangle
\end{equation}
In Equations~\ref{eq:7} and \ref{eq:8}, the symbols $\left[ ...\text{ },\text{ }... \right]$  and $\left\{ ...\text{ },\text{ }... \right\}$  denote the commutator and anticommutator, respectively, and $\left\langle ... \right\rangle$ represents the expectation value evaluated in the Heisenberg picture. 
Green functions are computed using the equation-of-motion method truncated at second order in projectile-surface coupling terms, following established procedures \cite{PhysRevB.93.195439,PhysRevB.96.075424,PhysRevB075422}. This approach yields fully time-resolved neutralization probabilities along projectile trajectories for each independent orbital channel under the infinite-$U$ approximation.

\subsection{Determination of the Hamiltonian parameters} \label{sec2c}
Surface wavefunctions are expressed as linear combinations of atomic orbitals (LCAO) centered on individual surface atoms:
\begin{equation}
{{\psi }_{\vec{k},\sigma }}=\sum\limits_{i,{{{\vec{R}}}_{S}}}{c_{i,{{{\vec{R}}}_{S}}}^{\vec{k},\sigma }\phi _{i,{{{\vec{R}}}_{S}}}^{\sigma }}
\end{equation}
where $\phi _{i,{{{\vec{R}}}_{S}}}^{\sigma }$ represents the wavefunction of atomic orbital $i$ centered at position ${{\vec{R}}_{S}}$. Expansion coefficients $c_{i,{{{\vec{R}}}_{S}}}^{\vec{k},\sigma }$ relate to the surface density matrix through: 
\begin{equation} \label{eq:10}
\rho _{i,{{\vec{R}}_{S}},j,{{\vec{R}}_{S'}}}^{\sigma }\left( \varepsilon  \right)=\sum\limits_{{\vec{k}}}{{{\left( c_{i,{{{\vec{R}}}_{S}}}^{\vec{k},\sigma } \right)}^{*}}c_{j,{{{\vec{R}}}_{S'}}}^{\vec{k},\sigma }\delta \left( \varepsilon -{{\varepsilon }_{\vec{k}\sigma }} \right)}
\end{equation}

The density matrix of the MoS$_2$ surface was calculated using the FIREBALL code within the LCAO-DFT approximation, with the LDA functional and Troullier-Martins pseudopotentials. Cutoff radii of 3.9, 4.5, and 5.0 a.u. were employed for the \textit{s}, \textit{p}, and \textit{d} orbitals of sulfur, and 5.0, 4.5, and 4.8 a.u. for those of molybdenum. The surface was relaxed using a 5×5 cell of four layers (300 atoms) for MoS$_2$. A sampling of 16 k-points in the first Brillouin zone was used. The lattice parameter obtained  was 3.20 \AA, in good agreement with the experimental value of 3.15 \AA{} (Ref.~[\onlinecite{PhysRevB.12.659}]). The density matrix was obtained from the DFT Hamiltonian via 
\[{{\rho }_{mn}}\left( \varepsilon  \right)=\frac{1}{\pi }\operatorname{Im}\left[ I\left( \varepsilon \pm i\eta  \right)-H \right]_{mn}^{-1}\]
where m and n summarize atom and state indices ($i,{{\vec{R}}_{S}}$ and $j,{{\vec{R}}_{S'}}$, respectively as in Eq.~\ref{eq:10}), $\varepsilon $ is the energy, $I$ is the identity matrix, and $\eta $ is a small value that provides a broadening to the atomic states, avoiding mathematical problems with zeros in the denominator.

Within this framework, coupling terms $V_{\vec{k}\alpha }^{\sigma }$ from Eq.~\ref{eq:4} decompose into superpositions of dimeric atomic hopping integrals $V_{\alpha ,i,{{{\vec{R}}}_{S}}}^{\sigma }$, 
\begin{equation}
V_{\vec{k}\alpha }^{\sigma }=\sum\limits_{i,{{{\vec{R}}}_{S}}}{c_{i,{{{\vec{R}}}_{S}}}^{\vec{k},\sigma }V_{\alpha ,i,{{{\vec{R}}}_{S}}}^{\sigma }}
\end{equation}
The bond-pair model \cite{PhysRevB.58.5007} is employed to compute coupling terms $V_{\alpha ,i,{{{\vec{R}}}_{S}}}^{\sigma }$ and energy levels ${{\varepsilon }_{\alpha }}$, excluding charge exchange between projectile and surface. Accurate parameter evaluation requires appropriate atomic basis sets for calculating one- and two-electron integrals. The one-electron hopping term incorporates both one- and two-electron contributions through mean-field approximation combined with many-body Hamiltonian overlap expansion.

A significant advancement in this work involves employing an extended Huzinaga basis set \cite{huzinaga1984gaussian} for the lithium projectile, specifically we use: (43333/433/43) basis including polarization \textit{p}  states, for Mo; (433/43) basis including polarization \textit{d} states, for S; and extended (73) basis enriched with 2\textit{p} orbital \cite{Huzinaga2p} for Li atoms. 
This expanded lithium basis represents a notable improvement over previous theoretical treatments \cite{1010630283586} that utilized a more restricted (43) basis omitting excited 2\textit{p} states. Incorporating these states is essential for rigorous evaluation of charge fluctuations into higher-energy channels.

The hybridization width, representing the imaginary component of the Green function self-energy, quantifies coupling strength between Li levels and surface states. Derived from Fourier transformation of Eq.~\ref{eq:8}, this width is given by:
\begin{equation} \label{eq:12}
{{\Gamma }^{\sigma }_{\alpha }}\left( {{\varepsilon }_{\alpha }} \right)=\pi \sum\limits_{i,{{\vec{R}}_{S}},j,{{\vec{R}}_{S'}}}{V_{\alpha ,i,{{{\vec{R}}}_{S}}}^{\sigma }V_{\alpha ,j,{{{\vec{R}}}_{S'}}}^{\sigma }}\rho _{i,{{\vec{R}}_{S}},j,{{\vec{R}}_{S'}}}^{\sigma }\left( {{\varepsilon }_{\alpha }} \right)\left[ 1+{{f}_{<}}\left( {{\varepsilon }_{\alpha }} \right) \right]
\end{equation}
where ${{f}_{<}}\left( {{\varepsilon }_{\alpha }} \right)$ denotes the Fermi function evaluated at projectile energy ${{\varepsilon }_{\alpha }}$. Equation~\ref{eq:12} demonstrates the dependence of hybridization width on the number of surface atoms effectively interacting with the projectile. Convergence studies indicate that including 13 atoms (7 Mo + 6 S) suffices for energy level and hybridization width convergence when Mo serves as scattering center \cite{1010630283586}.
The energy-resolved surface density matrix previously calculated for MoS$_2$ is employed to describe local electronic structure entering Eq.~\ref{eq:10}. These matrix elements derive from first-principles Density Functional Theory calculations using the FIREBALL code \cite{pssb.201147259}. While convergence of energy levels and widths is achieved with 13-atom clusters, the density matrix inherently incorporates information about extended surface character beyond cluster boundaries.

The integro-differential equations resulting from the equation-of-motion method of Eq.~\ref{eq:7} and Eq.~\ref{eq:8} were solved numerically using a time step of $\Delta t = 0.1 \, \text{a.u.}$ and an energy step of $\Delta \varepsilon = 0.05 \, \text{eV}$, which ensure convergence of the final occupations.

\subsection{Trajectory modeling and collision geometry}
\begin{figure}
\includegraphics[width=\columnwidth]{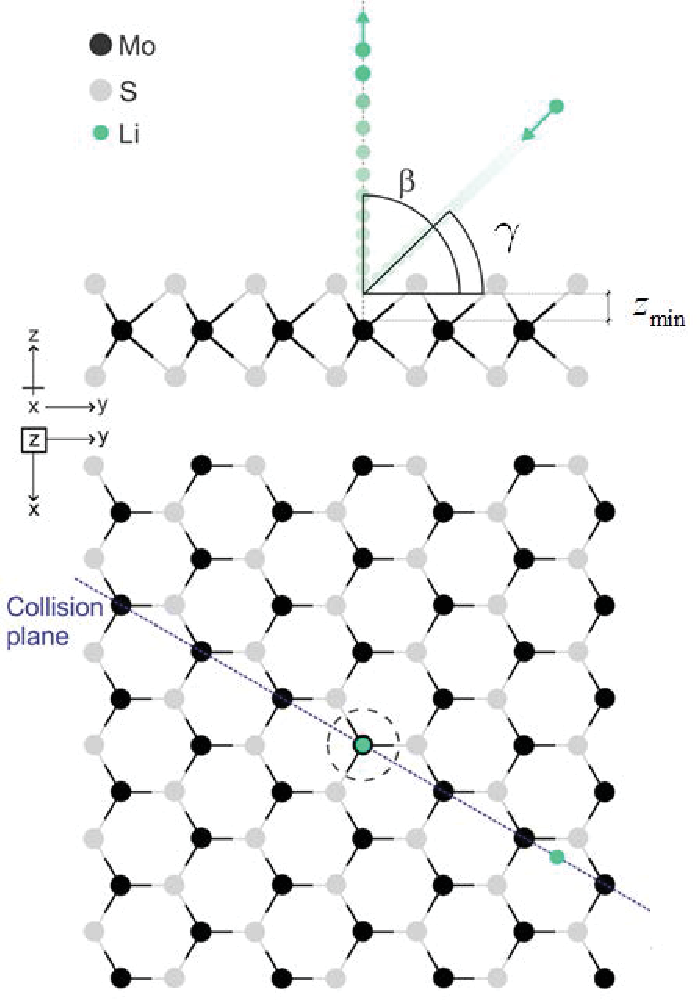}%
\vspace{-10pt}
\caption{\label{fig1} Geometric configuration scheme of the collisional system used for the
calculation. In the top panel (side view), the incident angle $\gamma$, exit angle $\beta$ are indicated, along with the return point ($z_{min}$), corresponding to the minimum distance between the Li projectile and the Mo scattering center. Mo, S, and Li atoms are represented by black, gray, and green spheres, respectively.}
\end{figure}

The target system comprises a MoS$_2$ trilayer in its 2H phase. This material exhibits a layered sandwich structure with a central Mo plane trigonally coordinated between two hexagonal S planes (see Fig.~\ref{fig1}). This bulk MoS$_2$ surfaces do not exhibit intrinsic surface or image states within the band gap, owing to their van der Waals nature and the absence of dangling bonds \cite{doi:10.1021/acsnano.6b01742}. Although such in-gap states can arise in few-layer MoS$_2$ due to enhanced surface effects and symmetry breaking, they are unlikely in the present system.

The complete details of the experimental setup, including surface preparation, TOF acquisition parameters, and the integration procedure used to extract the charge fractions, are described in Ref.~[\onlinecite{1010630283586}]. Briefly, the experiments were performed under UHV conditions using a pulsed beam of Li$^+$ in the energy range of 2.5--8.0\,keV, with a backscattering geometry of $\gamma / \beta=45^\circ /90^\circ $. The elastic peak corresponding to scattering with Mo was isolated by integration over time windows, while the S peak could not be fully resolved due to its overlap with multiple scattering contributions from Mo.

Collisions are modeled as head-on binary events where projectiles follow straight-line trajectories normal to the surface, approaching Mo scattering centers, reaching turning points, and retracing their paths. The geometry corresponds to incidence angle $\gamma=45^\circ$ and exit angle $\beta=90^\circ$ relative to the surface, matching experimental backscattering conditions \cite{1010630283586}. This configuration minimizes theoretical uncertainties since perpendicular exit trajectories ($v_\parallel=0$ during departure) eliminate parallel velocity effects that receive only approximate treatment in our model.

The exit kinetic energy ($E_{out}$) of the projectile, resulting from a binary elastic collision with the scatter atom is calculated as $E_{out} = \lambda E_{in}$, where $\lambda$ represents the kinetic energy loss factor \cite{oura2013surface}. This factor is equal to 0.78 for Li-Mo systems under specified scattering geometry. Consequently, exit velocities are reduced relative to entrance velocities ($v_{out}=\sqrt{\lambda}v_{in}$), asymmetrically modifying time evolution rates of Hamiltonian parameters between incoming and outgoing branches. The velocities along the incoming and outgoing branches are both taken equal to the perpendicular component of the experimental velocity vector.

Hamiltonian parameters are evaluated along straight trajectories as functions of perpendicular distance from the surface, $|z\left(t\right)|$, with return points (distance of closest approach) fixed at $z_{min}=1$ a.u. from Mo scattering centers for all incident energies examined. While this value somewhat exceeds predictions from ZBL potentials \cite{Ziegler} (the ZBL potential predicts return points ranging from 
$0.41$ a.u.\ at $2.5$ keV to $0.22$ a.u.\ at $8$ keV), previous investigations demonstrate weak sensitivity of neutralization probabilities to this parameter \cite{PhysRevB.100.085432,PhysRevB.102.115406,PhysRevA.107.032803}.

The time dependence of the Hamiltonian parameters arises through the projectile-surface distance $z\left( t \right)$, which is determined by the classical trajectory. At each point along this trajectory, the energy levels and couplings are evaluated using the \textit{ab initio} methods described in Sec. \ref{sec2c}. Thus, the Keldysh Green functions are solved along a time-dependent path that maps directly onto the experimental geometry, ensuring a realistic description of the charge exchange dynamics

\subsection{Model justification, Multi-channel treatment and correlation approximation}
The infinite-$U$ approximation is well-justified by the experimentally observed low yield of negative ions ($<$ 5\%) \cite{1010630283586}. Although this approximation originates from neglecting doubly occupied configurations, it can be reinterpreted as an independent-channel approximation that effectively incorporates spin fluctuations. From this standpoint, the negative ion fraction can even be derived within the same independent-channel framework by working in the hole representation. This approximation has proven successful in similar systems involving alkali projectiles \cite{PhysRevA032901,PhysRevB.96.075424,PhysRevB075422,jp4116673,GARCIA2009597}   

A central feature of this work involves independent treatment of each orbital channel. Charge dynamics for 2\textit{s}, 2\textit{$p_x$}, 2\textit{p$_y$}, and 2\textit{p$_z$} channels are solved separately using the formalism described above, yielding independent neutralization probabilities $n_{\alpha }^{0}\left( t \right)$ along trajectories. These independent calculations do not account for competition between different orbitals of the same atom. 
To achieve physically realistic descriptions, electronic correlation between channels is introduced approximately at final states. Correlation-corrected neutralization probabilities for each channel $P_{\alpha }^{0}$  and total neutralization $P_{total}^{0}$ are computed using: 
\begin{equation} \label{eq:13}
P_{\alpha }^{0}=n_{\alpha }^{0}\prod\limits_{\mu \ne \alpha }{\left( 1-n_{\mu }^{0} \right)}
\end{equation}
And
\begin{equation} \label{eq:14}
P_{total}^{0}=\sum\limits_{\alpha }{P_{\alpha }^{0}}
\end{equation}

The product of probabilities correction expressed in Eqs. \ref{eq:13} and \ref{eq:14} imposes electron exclusion among the considered channels in the final state of the projectile. Since each orbital channel is calculated independently, the occupations $n_{\alpha }^{0}$ represent neutralization probabilities as if each channel were the only one available. However, in reality, the projectile can host at most one electron. To correct this, we project the independent probabilities onto the physical subspace of single occupation. The construction assumes that, in the final state, the occupation events in different channels are statistically independent, but the exclusion condition is imposed a posteriori. Thus, the probability that channel $\alpha $ is occupied and all other channels $\mu \ne \alpha $ are empty is the product of the probability that channel $\alpha $ is occupied, $n_{\alpha }^{0}$, and the probability that each other channel is empty, $\left( 1-n_{\mu }^{0} \right)$. This factorization is analogous to the exclusion rule for independent fermions and has been used in previous works of our group to estimate correlation effects when a full dynamical treatment is computationally prohibitive~\cite{PhysRevB.96.075424}. We acknowledge that this correction is applied only to the final state and does not capture the dynamical correlation along the trajectory, where instantaneous occupations should be coupled.

This approximation estimates electronic correlation effects using individually calculated final occupations. Although this approach captures correlation only at final states rather than dynamically throughout trajectories, it provides first-order estimates of inter-channel competition importance and excited state roles in overall neutralization processes.

The product of probabilities correction described above can be viewed as a mean-field approximation for the orbital occupations, where we replace the expectation value of a product of occupations by the product of their individual expectation values: $\left\langle {{{\hat{n}}}_{\alpha }}\left( 1-{{{\hat{n}}}_{\mu }} \right) \right\rangle \approx \left\langle {{{\hat{n}}}_{\alpha }} \right\rangle \left\langle \left( 1-{{{\hat{n}}}_{\mu }} \right) \right\rangle =n_{\alpha }^{0}\left( 1-n_{\mu }^{0} \right)$. This neglects covariances between channel occupations, which is justified as a first-order approximation when the occupation of each channel is relatively small, as is the case here.

\section{Results and Discussion} \label{sec:3}
\subsection{Variation of the 2\textit{s} Level due to the Inclusion of the 2\textit{p} Basis Set}

\begin{figure}
\includegraphics[width=\columnwidth]{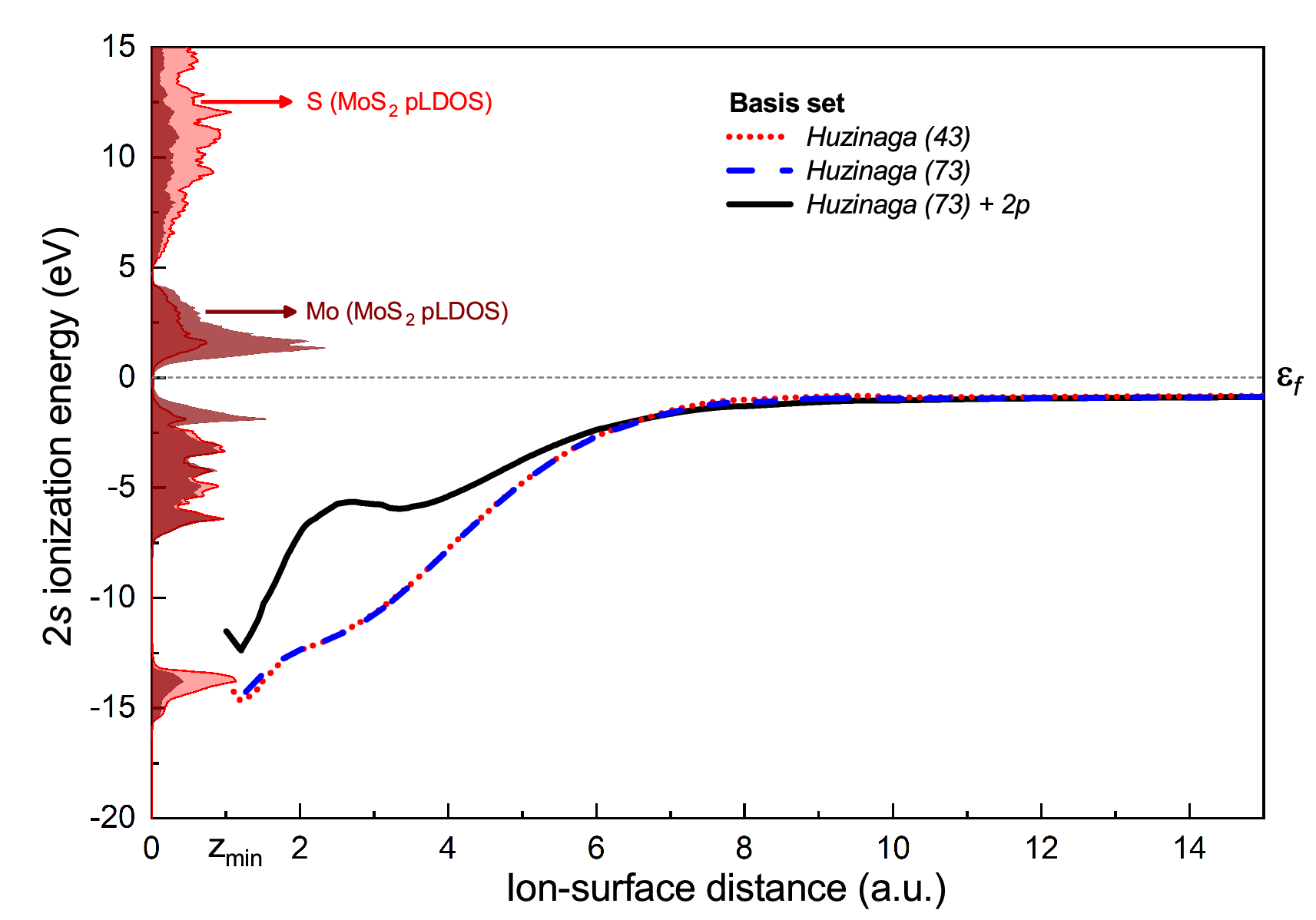}%
\vspace{-10pt}
\caption{\label{fig2} Ionization energy of the Li 2\textit{s} orbital as a function of the projectile$-$surface perpendicular distance, calculated with three different basis set. The dotted line corresponds to the Huzinaga 43 basis set (previously reported in Ref.~[\onlinecite{1010630283586}] and included here for comparison), while the dashed and solid lines correspond to the new Huzinaga 73 basis set and the Huzinaga 73 basis set augmented with 2\textit{p} empty orbitals \cite{Huzinaga2p}, respectively. On the left, the dark and light red shaded areas represent the local density of states of MoS$_2$ projected onto the Mo and S atoms, respectively. All energies are referenced to the Fermi level (${{\varepsilon }_{f}}$) of the surface (horizontal grey dashed line), assuming a work function of 4.54 eV for MoS$_2$ \cite{doi:10.1021/acsnano.6b01742}. The calculations extend up to the return point ($z_{min}$), set at 1 a.u.}

\end{figure}

In Fig.~\ref{fig2} we present the ionization energy of the Li 2\textit{s}  orbital (${{\varepsilon }^{io}_{2s}}=E_{1}^{2s}-E_{0}^{2s}$) as a function of the projectile$-$surface distance, for a Li atom interacting with a Mo atom and twelve neighboring atoms on a MoS$_2$ surface and calculated with three different basis set. 

This Figure demonstrates the impact of the atomic basis set selection on the calculated parameters. A comparison is presented between the ionization energy of the 2\textit{s} level obtained with two basis sets that only include 1\textit{s} and 2\textit{s} orbitals (Huzinaga 43 and 73), and the extended basis set (which includes the 2\textit{p} basis \cite{Huzinaga2p}) employed in this work.
The inclusion of 2\textit{p} orbitals in the lithium basis is found to substantially alter the position of the 2\textit{s} level, particularly within the strong-interaction region ($1 < z < 4$ a.u.). The most significant deviation occurs at approximately $z \approx  3$ a.u., where the projectile is closest to the sulfur atoms of the top layer ($z = 0$ corresponds to the Mo scattering center).
The extended basis set results in a shallower, more realistic potential well. This improvement is attributed to the symmetric orthogonalization and the second-order overlap (now includes the 2\textit{p} orbitals) expansion employed in the bond-pair model \cite{PhysRevB.58.5007}. This refinement in the projectile's electronic structure directly influences the efficiency of resonant charge transfer processes and, as discussed below, substantially enhances the predictive capability for final neutral fractions.

\subsection{Energy levels evolution and its hybridization width}
\begin{figure*}
\includegraphics[scale=0.6]{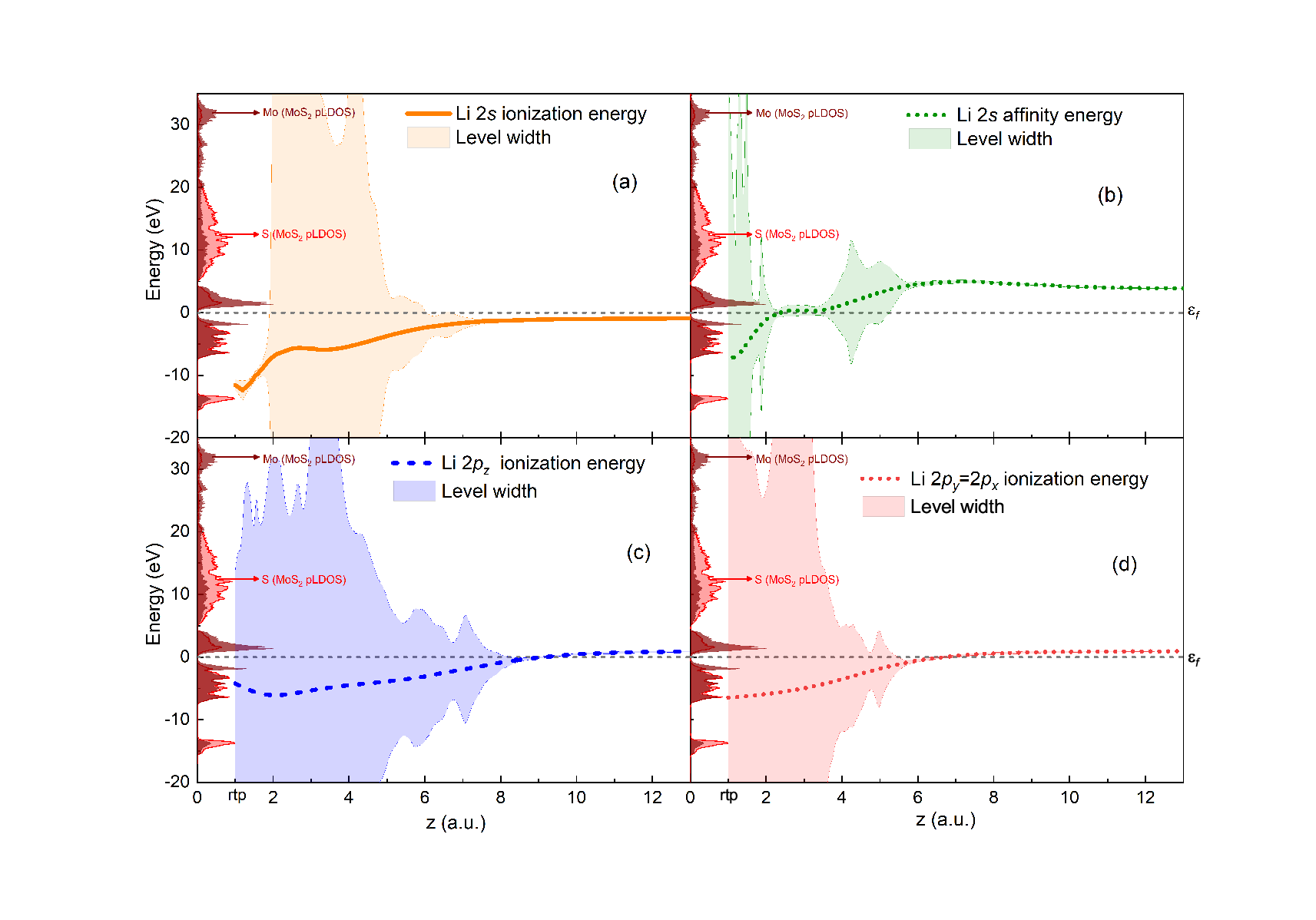}%
\vspace{-40pt}
\caption{\label{fig3} Variation of the atomic energy levels $\varepsilon_\alpha$ (lines) and the corresponding hybridization widths ${\Gamma }^{\sigma }_{\alpha }$ (shaded areas) as a function of the ion-surface distance $z$ for Li atoms colliding head-on with Mo atoms of the surface. Calculations were performed using the extended basis set. The panels display the distinct orbital channels: (a) Li $2s$ ionization energy, (b) Li $2s$ affinity energy, (c) Li $2p_z$ ionization energy, and (d) Li $2p_{x,y}$ ionization energy. In all panels, the projected local density of states (pLDOS) of MoS$_2$ on the Mo and S atoms is shown on the left vertical axis, allowing a direct visual correlation between the projectile's energy levels and the electronic band structure of the substrate.}
\end{figure*}

Figure~\ref{fig3} shows the variation of atomic energy levels $\varepsilon _{\alpha}$ and corresponding hybridization widths ${\Gamma }^{\sigma }_{\alpha }$  (as error bar: shadowed area) with ion-surface distance $z$ for Li atoms colliding frontally with topmost-layer Mo atoms. All calculations were performed using the extended basis set. The left panels shows MoS$_2$ local density of states projected onto Mo and S atoms, enabling direct visual correlation between projectile energy levels and substrate electronic band structure.

Energy level behavior divides into two distinct spatial regimes. At large separations ($z>$12 a.u.), atomic levels approach their asymptotic vacuum values relative to the surface Fermi level (Image potential is neglected, as MoS$_2$ lacks metallic character). These asymptotic values correspond to free atomic energies: Li ionization energy ($\varepsilon _{2s}^{io}=E_{1}^{2s}-E_{0}^{2s}= -$5.39 eV relative to vacuum), affinity energy ($\varepsilon _{2s}^{aff}=E_{2}^{2s}-E_{1}^{2s}= -$0.62 eV), and excited 2\textit{p} state energy (${\varepsilon_{2p}} = E_{1}^{2p} - E_{0}^{2p} = -3.54$ eV) \cite{10.1063/1.1800011}. In Fig.~\ref{fig3} these values are referenced to the Fermi level accounting for MoS$_2$ work function ($\Phi =$4.54 eV \cite{doi:10.1021/acsnano.6b01742}). 

All levels undergo abrupt downward shifts resulting from strong electrostatic interactions between projectile electrons and substrate atoms. 

Concurrently with energy variation, level broadening occurs due to coupling with the continuum states of the solid [Eq.~\ref{eq:12}]. Figure~\ref{fig3} illustrates the orbital dependence of this coupling. For the 2\textit{s} orbital, the ionization energy (Fig.~\ref{fig3}a) remains predominantly below the Fermi energy. Hybridization becomes significant when this energy resonates with the valence band density of states (DOS), primarily associated with Mo $d$ and S $p$ orbitals. This behavior of the level and its width indicates that charge capture processes should dominate over loss processes, making this the predominant neutralization channel.

The affinity level (Fig.~\ref{fig3}b) remains above the Fermi energy, except at very short distances ($z <$ 2.5 a.u.), where it resonates with the surface band gap. This results in a sharp reduction in the broadening (for 2.3 a.u. $< z <$ 3.5 a.u.), with appreciable width only at very close distances, indicating that electron loss processes will dominate. Consequently, the probability of negative ion formation is expected to be negligible.

Regarding the 2\textit{p} orbital, the perpendicularly oriented 2\textit{p$_z$} orbital (Fig.~\ref{fig3}c)  exhibits greater spatial extension toward the solid, leading to earlier and more pronounced hybridization compared to the planar 2\textit{p$_{x,y}$} orbitals (Fig.~\ref{fig3}d). These levels cross the Fermi energy while simultaneously acquiring a finite hybridization width, opening resonant ionization channels.

For all levels, at very short distances, strong interaction with the solid causes the atomic levels to lose their discrete character and broaden substantially, transforming into wide resonances. Although the central ionization energy of the 2\textit{s} orbital lies nominally below the Fermi level the large hybridization width causes substantial portions of the spectral distribution to extend above ${{\varepsilon }_{f}}$. This generates overlap with unoccupied MoS$_2$ conduction band states, enabling electron return to the solid (substrate neutralization or projectile reionization) via tunneling. 

The behavior of the hybridization widths at short distances also provides a physical justification for the weak sensitivity of the calculated charge fractions to the precise value of the return point $z_{min}$. As seen in Fig.~\ref{fig3}, at close projectile-surface separations each channel falls into one of two limiting regimes: either the hybridization width $\Gamma^{\sigma}_{\alpha}$ nearly vanishes because the level resonates with the band gap (ionization $2s$ channel, panel a), so that the projectile effectively decouples from the surface and the charge dynamics become insensitive to the exact turning point; or $\Gamma^{\sigma}_{\alpha}$ spans the entire surface bandwidth (affinity $2s$ and ionization $2p$ channels, panels b, c, and d), so that electron capture and loss processes mutually compensate and further reduction of $z_{min}$ does not alter the net charge transfer. In both limits, the asymptotic charge fractions are robust against variations in $z_{min}$, consistent with the findings of Refs.~[\onlinecite{PhysRevB.100.085432,PhysRevB.102.115406,PhysRevA.107.032803}].

\subsection{Independent and correlated neutralization channels}
\begin{figure}
\includegraphics[width=\columnwidth]{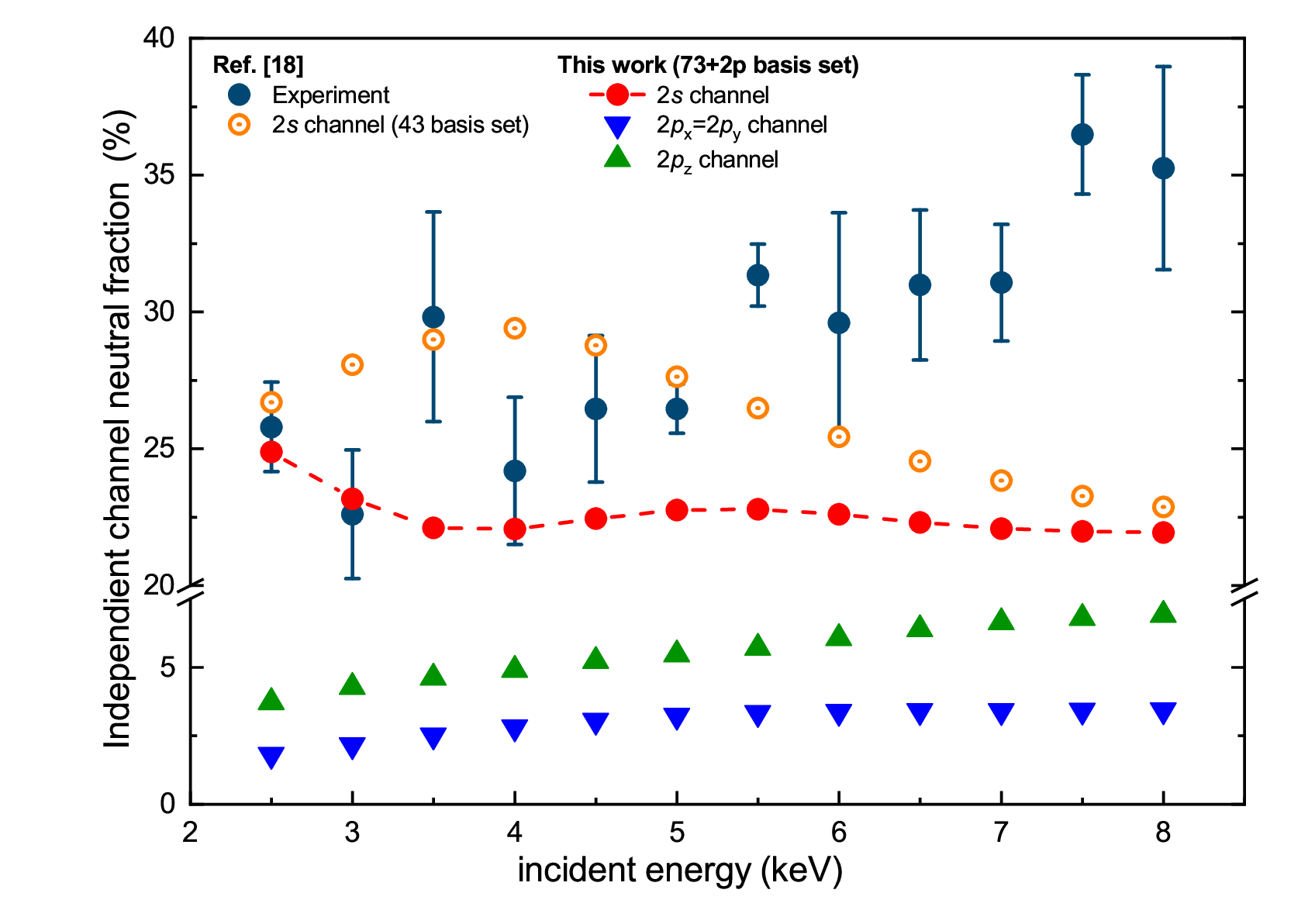}
\vspace{-20pt}
\caption{\label{fig4} Independent channel neutral fractions as a function of incident energy (in keV). Experimental data from Ref.~[\onlinecite{1010630283586}] (filled circles with error bars) are compared with the calculations of this work using the extended basis set ($73 + 2p$). The individual contributions of the $2s$ channel (filled circles with dashed lines), the $2p_z$ channel (up triangle), and the $2p_{x,y}$ channels (down triangle) are shown. The previous calculation for the $2s$ channel with the restricted 43 basis set (open circles) is also included.}
\end{figure}

Figure~\ref{fig4} presents the neutral fractions for independent channels as a function of the incident energy, comparing the experimental data from Ref.~[\onlinecite{1010630283586}] with the calculations performed using the restricted basis set (43) \cite{1010630283586} and the extended basis set that includes the 2\textit{p} orbital (present work).

The most notable improvement is the better agreement with the experimental trend for the 2\textit{s} channel relative to previous calculations, which employed simpler atomic basis for lithium. This enhancement arises directly from the inclusion of 2\textit{p} orbitals in the lithium basis set, which modifies the 2\textit{s} energy levels (see Fig.~\ref{fig2}) and refines the dynamical evolution to more closely match experimental behavior, particularly at low energies. At high energies ($>$7 keV), the differences between simple and extended basis diminish, as kinematic broadening (uncertainty associated with velocity) dominates over static electronic structure details.

Examination of 2\textit{p} orbitals reveals equivalent neutral fractions for 2\textit{p$_x$} and 2\textit{p$_y$} channel due to rotational symmetry, showing energy-dependent growth saturating near 5 keV. The 2\textit{p$_z$} channel exhibits nearly linear increase across the entire range. The 2\textit{p} state contributions are significantly smaller than the fundamental 2\textit{s} contribution, as expected given the latter's energetic favorability for electron capture.

\begin{figure}
\includegraphics[width=\columnwidth]{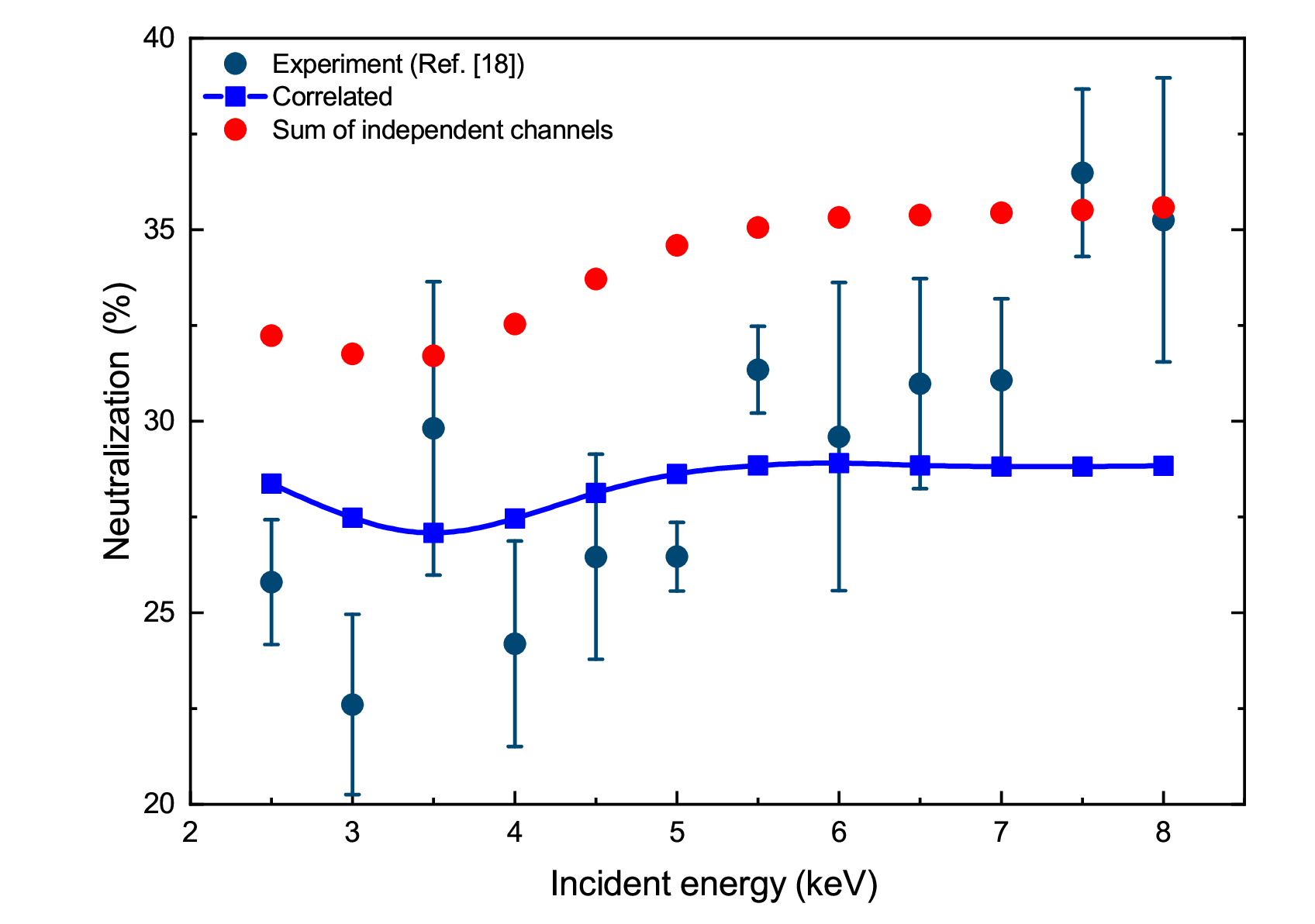}%
\vspace{-10pt}
\caption{\label{fig5} Total neutral fraction compared with experimental results from Ref.~[\onlinecite{1010630283586}] (filled circles with error bars). Two theoretical approaches are shown: the sum of the independent channels without correlation (filled circles) and the model incorporating electronic correlation between channels (filled squares with solid line).}
\end{figure}

Figure~\ref{fig5} presents the total neutral fraction in comparison with the experimental results of Ref.~[\onlinecite{1010630283586}]. Two different approaches were used to calculate this neutral fraction: one that neglects electronic correlation between channels (i.e., the sum of the independent channels) and another that incorporates such correlation, derived from the independent channel probabilities as detailed in Eqs.~\ref{eq:13} and \ref{eq:14}. The independent-channel approximation overestimates the experimental data across nearly the entire energy range studied, while the approximation that includes correlation yields a significantly better representation of the experimental values. The inclusion of correlation effects improves global agreement with experiment, bringing the theoretical values closer to the measured data, particularly along the high-energy slopes when compared with the single 2\textit{s} channel (see Fig.~\ref{fig4}).

However, two important limitations of this approximation warrant mention. First, the probabilistic correction is applied only to asymptotic final values (i.e., in the limit  $t\to \infty $), whereas physically, correlation acts dynamically throughout the trajectory. If the 2\textit{s} channels become occupied during the collision, they should instantaneously block charging in the 2\textit{p} channel (and vice versa), thereby modifying the charge history beyond just the final outcome. Second, a rigorous inclusion of correlation would require extended Hamiltonian that explicitly couple all orbital channels through surface band interactions and satisfy strict global normalization conditions $\left( \sum\limits_{\alpha ,\sigma }{{{n}_{\alpha ,\sigma }}\le 1} \right)$ \cite{PhysRevB.96.075424}. Under the assumption of infinite coulomb repulsion ($U\to \infty $), the projectile captures only a single electron (either in the 2\textit{s} or 2\textit{p} orbitals), which precludes negative ion formation. Dynamically, the effective hybridization width of each channel becomes modified by the couplings of the remaining channels. Solving this exact scheme would demand the simultaneous integration of at least eight coupled integro$-$differential equations (for Green's functions), incurring prohibitively high computational costs.

\subsection{Negative ion formation analysis via hole formalism}
Thus far, analysis has concentrated on neutralization processes (Li$^{+}\leftrightarrow $ Li$^{0}$), assuming strong Coulomb repulsion prevents simultaneous double electron occupation in projectile orbitals. However, complete charge dynamics description requires evaluation of negative ion formation (Li$^{-}$) probabilities, representing second electron capture (in 2\textit{s} orbital) by neutral atoms.

To address this calculation while maintaining consistency with Green’s function formalism and the   $U\to \infty$ approximation, a canonical Hamiltonian transformation to hole representation is employed \cite{PhysRevB.93.195439}. This strategy inverts definitions of 'particle' and 'vacuum'. In this transformed scheme, states doubly occupied by electrons (Li$^{-}$ ions) become "hole vacuum" states, while neutral states (Li$^{0}$) correspond to single hole presence. Applying infinite correlation approximation ($U_{hole}\to \infty$) physically prohibits "double hole occupation" (corresponding to original Li$^{+}$ states), restricting dynamics to transitions between Li$^{-}$ and Li$^{0}$.

Applying this transformation to the finite-$U$ Hamiltonian \cite{PhysRevB.80.235427} and taking the limit $U\to \infty$ , we obtain exactly the same structure as the Hamiltonian giving by Eqs.~\ref{eq:2}, \ref{eq:3} and \ref{eq:4}, but with parameters redefined as follows:  ${{\varepsilon }_{{\vec{k}}}}$ is replaced by  $-{{\varepsilon }_{{\vec{k}}}}$, $E_{0}^{2\textit{s}}$ is replaced by $E_{2}^{\textit{2s}}$ (total energy with two electrons in the 2\textit{}s orbital),  $E_{1}^{2\textit{s}}$ remains $E_{1}^{2\textit{s}}$, and $V_{\vec{k}}$ is replaced by  $-V_{\vec{k}}$. Since the equations of motion depend only on the commutation relations of the operators, and these are maintained in both representations (electrons and holes), we can use the same code to calculate the probabilities of negative ion formation, taking into account the aforementioned changes in parameters. These changes lead us to redefine the orbital energy as  $E_{2}^{2\textit{s}}-E_{1}^{2\textit{s}}$, which is associated with the affinity level of the 2\textit{s} orbital (see Fig.~\ref{fig3}b).
Within this hole scheme, the code calculates the probability of having a single hole ($n_{h}^{2s}$), i.e., of being neutral. By the normalization condition in the hole representation, the independent probability of observing the hole vacuum state (which is equivalent to the negative ion doubly occupied by electrons) is obtained directly as:
\begin{equation}
n_{\text{Li}^-}^0 = 1 - \sum_\sigma n_{h,\sigma}^{2s}
\end{equation}

\begin{figure}
\includegraphics[width=\columnwidth]{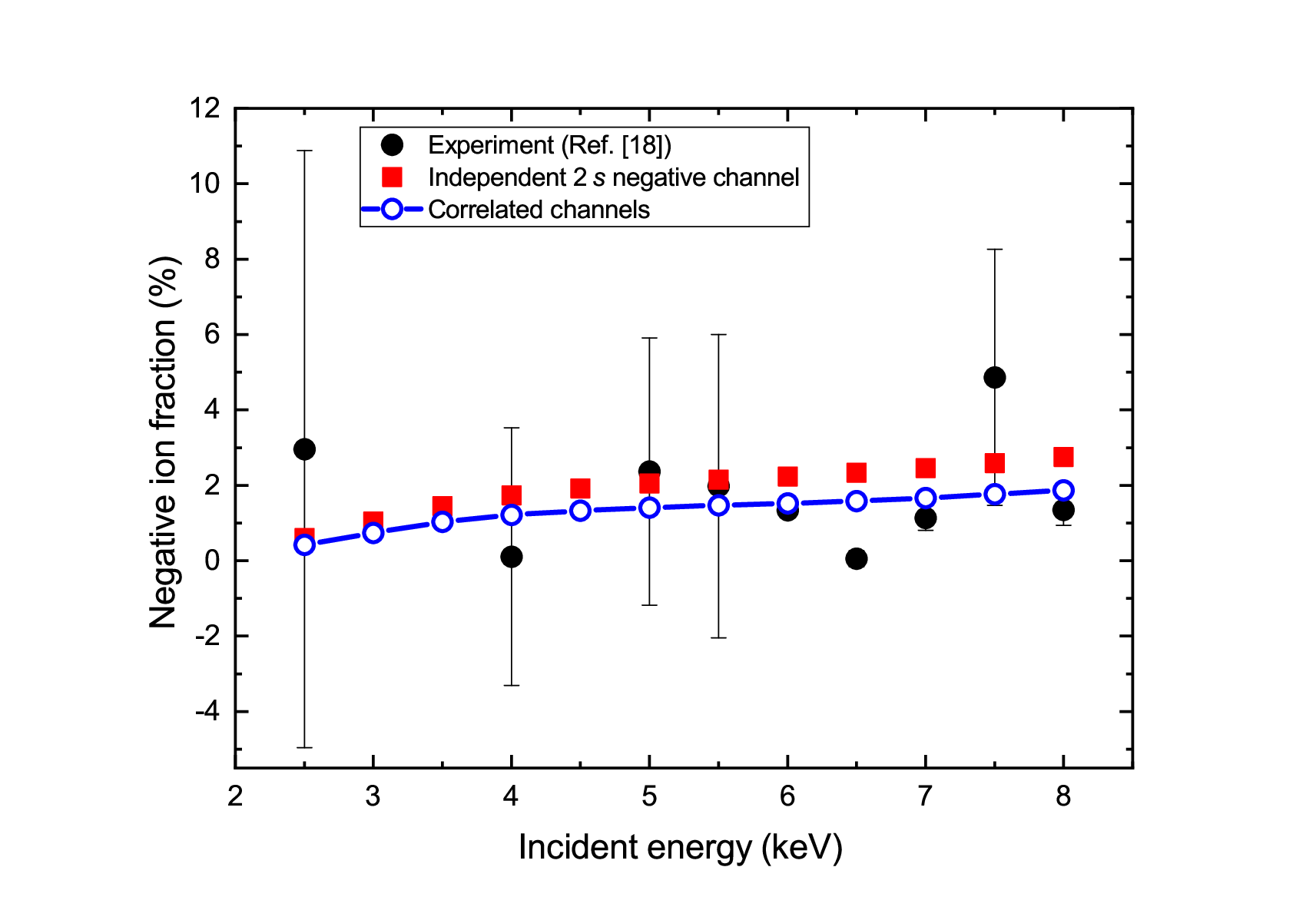}%
\vspace{-20pt}
\caption{\label{fig6} Negative ion fraction as a function of incident energy (in keV). Experimental results from Ref.~[\onlinecite{1010630283586}] (filled circles with error bars) are compared with calculations for the independent $2s$ negative channel (filled squares) and the correlated channels (open circles with solid line).}
\end{figure}

Figure~\ref{fig6} presents a detailed comparison between the calculated negative ion fractions  and the available experimental data from Ref.~[\onlinecite{1010630283586}]. Here, as before, we present two approximations for the negative ion fractions: one that neglects correlation between channels and one that includes it. The approximation neglecting correlation corresponds simply to the newly described independent channel result ($n_{\text{Li}^-}^0$).
On the other hand, the probability of negative ions that includes correlation is calculated in a similar way to Eq.~\ref{eq:13}, as follows.
\begin{equation}
P_{\text{Li}^-} = n_{\text{Li}^-}^0 (1 - n_{2s}^0) (1 - n_{2p_z}^0) (1 - n_{2p_x}^0) (1 - n_{2p_y}^0)
\end{equation}
Again, the calculation that accounts for correlation between channels provides the best agreement with the experimental data.

Experimental measurements for this ion fraction present considerable technical difficulties, reflected in substantial error bars due to significant detection uncertainties for low ionic currents, as well as oscillatory fluctuations where the experimental points lack smooth behavior and exhibit abrupt jumps between consecutive energies that obscure clear trend identification. Despite the scatter in the experimental data, the theoretical calculations satisfactorily reproduce the order of magnitude of the phenomenon. Our results predict fractions between 0.5\% and 2\%, values that fall within the experimental uncertainty ranges and follow the average trends of the data.

Low negative ion production rates find direct physical explanation in system electronic structure. Lithium's electron affinity energy level resides energetically above MoS$_2$ surface Fermi level during most relevant interactions. Moreover, this level remains (asymptotically) above energies corresponding to 2\textit{p} orbitals. Consequently, transitions toward excited neutral configurations (2\textit{p} channel) are energetically more favorable and thus substantially more probable than second electron capture in 2\textit{s} channels for negative species formation.

This energy configuration renders second electron retention highly unfavorable. Although dynamical processes at short distances may permit transient second electron capture, neutralization rates (loss of extra electrons to empty conduction band states) dominate. Hybridization to negative channel (Fig.~\ref{fig3}b) thus act more efficiently to "empty" negative states than to fill them, preventing projectile departure from surfaces with net $-1$ charges.

Given this fraction's low magnitude, its influence on global neutralization results does not significantly alter charge balance. For precise quantification, negative channel contributions are incorporated within final-state electronic correlation approximations (Eq.~\ref{eq:13}). Specifically, $(1 - n_{Li^-}^0)$ terms are added to products, probabilistically conditioning systems and enabling strict quantification of only neutral states by discounting populations decaying to negative states. Including this correction yields final results practically indistinguishable from previously calculated neutralization (because the small negative ion formation), except at high incident energies where very small decreases appear (not shown).

In conclusion, negative channel inclusion through correlation schemes further approximates models to physical realities of multi-channel systems, refining high-energy observables without modifying dominant charge transfer dynamics.

We emphasize that the infinite-$U$ limit is well justified for the present system by the experimental observation that the negative ion fraction is below 5\%~\cite{1010630283586}, indicating that double occupation of the projectile is extremely unlikely. Our theoretical calculations of the negative ion fraction (Fig. 6) reproduce both the magnitude and trend of the experimental data, with values between 0.5\% and 2\%, further supporting the weak coupling between the 0-1 and 1-2 electron fluctuation channels. While a finite-$U$ treatment could in principle be implemented, the expansion in the projectile-surface coupling $V_{\vec{k}\alpha }^{\sigma }$ (required by the finite-$U$ formalism of Ref.~[\onlinecite{PhysRevB.80.235427}]) is not valid in this system because the hybridization widths $\Gamma _{\alpha }^{\sigma }$ are comparable to or larger than the effective $U$ in the region where charge transfer is most active (see Fig. \ref{fig3}). Numerical attempts to implement the finite-$U$ model yielded inconsistencies, including non-physical oscillations and convergence difficulties. We therefore adopt the well-justified infinite-$U$ limit as a computationally tractable approximation that captures the essential physics of the charge exchange process.

\subsection{Temporal evolution along trajectories}
\begin{figure}
\includegraphics[width=\columnwidth]{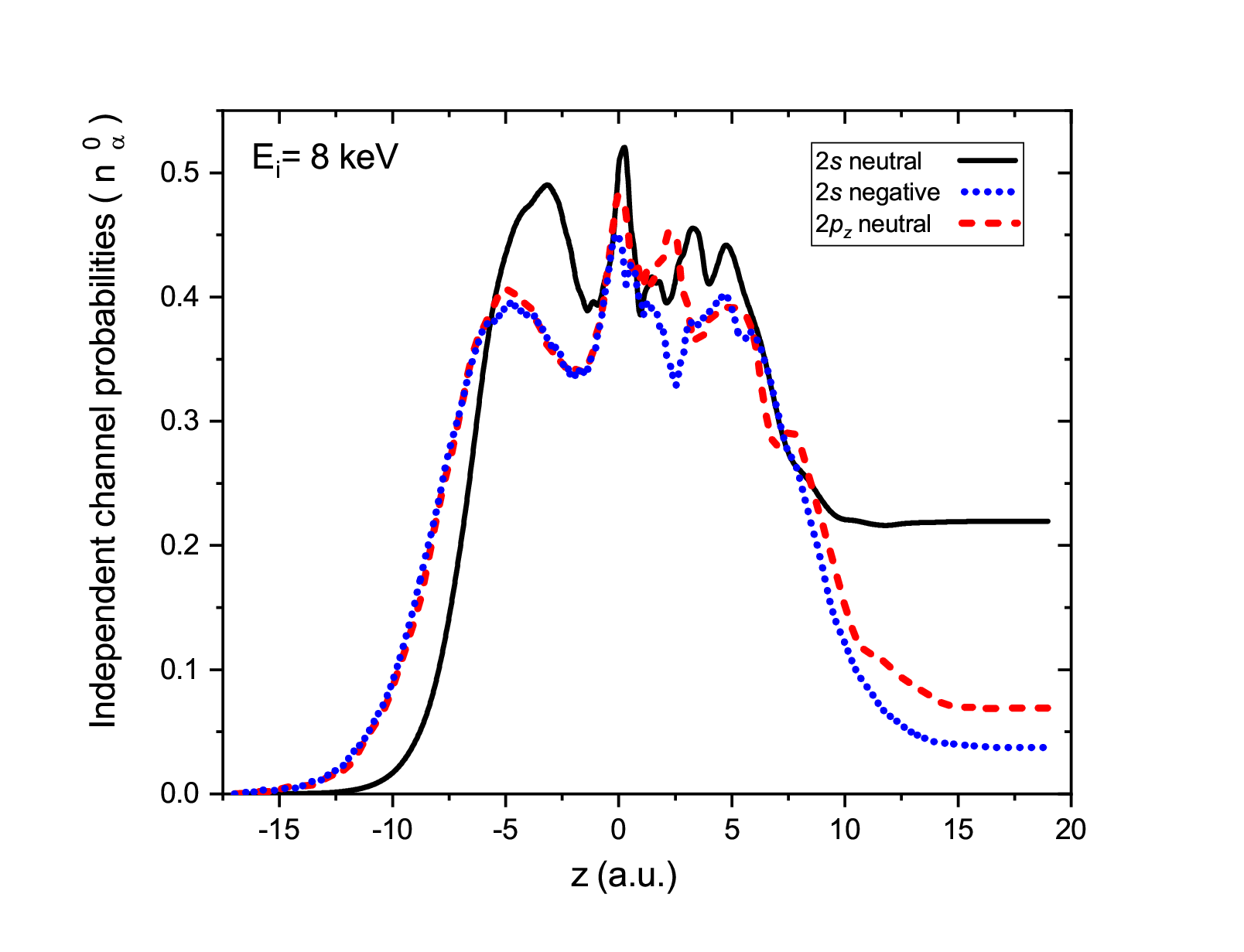}%
\vspace{-20pt}
\caption{\label{fig7} Spatial evolution of the independent channel probabilities ($n_\alpha^0$) along the projectile trajectory for an incident energy of $8$ keV. The distance axis is defined such that $z=0$ corresponds to the return point ($z_{min}=1$ a.u. from the surface). The neutral $2s$ (solid line) and $2p_z$ (dashed line) channels are plotted alongside the negative $2s$ channel (dotted line). It is evident that in the region of maximum proximity to the surface ($z=0$), the summed individual occupations exceed unity, demonstrating a violation of the norm and indicating the need for dynamic correlation at short distances.}
\end{figure}

To understand the limitation of applying the correlation approximation only to the final asymptotic values, mentioned above, it is necessary to examine the behavior of the charge state evolution. Figure~\ref{fig7} shows the instantaneous evolution of the channel occupancies within the independent-channel approximation ($n_{\alpha }^{0}\left( t \right)$) for channels 2\textit{s} (neutral and negative probabilities) and 2\textit{p$_z$} (neutral probabilities) along trajectories (incoming region $z<$ 0, outgoing region $z>$ 0) for an incident energy of 8 keV. The 2\textit{p$_x$} and 2\textit{p$_y$} channels exhibit similar behavior (not shown).

A remarkable feature emerges near return points ($z \approx  0$), where individual occupations of each channels approach or exceed 0.5. Linear summation of these occupations (including 2\textit{p$_x$} and 2\textit{p$_y$} contributions) would yield total occupation probabilities strictly exceeding unity $\left( \sum\limits_{\alpha }{{n_{\alpha  }^{0}} > 1} \right)$. 

This probabilistic excess does not represent real multiple charge state formation but rather a purely mathematical artifact arising from independent channel evaluation that ignores inter-channel correlation. This result reinforces the conclusion that independent channel treatments overestimate charge at short distances and indicates the necessity of dynamical multi-channel correlation treatments throughout the trajectories, not merely at the final states. 

This violation of the normalization condition arises because each channel is treated as if it were the only one available to the electron. Physically, when the $2s$ channel becomes populated, the $2p$ channels must be correspondingly depopulated. This dynamical coupling is not captured by our final-state correction, highlighting the need for a fully coupled multi-channel treatment.

A fully dynamical treatment of electronic correlation is expected to further improve the agreement with experimental data. Such an approach would naturally lead to an enhancement of the hybridization widths for each channel while enforcing a strict exclusion principle: as one channel becomes populated, the others must simultaneously deplete. Within this correlated framework (specifically between the 2\textit{s} and 2\textit{p} neutral channels) the Li 2\textit{s} channel would populate first, followed by the 2\textit{p} orbitals (starting with 2$p_z$ and then 2$p_{x,y}$) at shorter distances, triggering a corresponding decay in the 2$s$ population. Consequently, at close proximity to the surface, the 2$p$ channels would reach higher occupancy levels than predicted by the independent-channel model. This redistribution of charge toward the 2$p $ manifold at high velocities could provide a more comprehensive explanation for the observed experimental trends.

\section{Conclusions} \label{sec:4}

In this work, we present a purely theoretical examination of charge exchange dynamics during low energy Li$^+$ collisions with MoS$_2$ surfaces, concentrating on the role of excited states in neutralization processes. Employing time dependent resonant charge transfer models based on Anderson Hamiltonians under the infinite-$U$ limit, neutralization probabilities for the 2\textit{s}, 2\textit{p$_x$}, 2\textit{p$_y$}, and 2\textit{p$_z$} channels were calculated independently, with Hamiltonian parameters derived from extended Huzinaga basis sets explicitly including lithium's 2\textit{p} orbitals.

The choice of basis set proves to be significant, as extended Huzinaga basis that include 2\textit{p} orbitals substantially modify the calculated 2\textit{s} energy levels in the strong interaction region ($z \approx  3$ a.u.), demonstrating the importance of adequate basis sets for describing ion$-$surface interaction parameters. Including the 2\textit{p} channels in the theoretical description leads to enhanced agreement with experiment, improving correspondence with previously measured experimental neutral fractions, particularly at higher incident energies ($6-8$ keV) where single channel calculations underestimate neutralization probabilities. 

Correlation effects prove essential, since independent channel occupations exceed unity at short projectile$-$surface separations, highlighting the necessity of including electronic correlation in the description; while probabilistic final state correlation approximations provide first order estimates, they fail to capture dynamical correlation effects throughout the trajectories. While the current probabilistic treatment of correlation is a first-order approximation, it is sufficient to demonstrate that the inclusion of the Li 2\textit{p} manifold significantly improves the theoretical description compared to the single-channel 2\textit{s} approach, narrowing the existing discrepancies with experimental data without the need for more computationally expensive dynamical multi-channel models.

Additionally, negative ion formation is found to be limited. Application of the hole formalism confirms that Li$^-$ formation probabilities remain below 2\% across the entire energy range examined, consistent with experimental observations.

Taken together, these results establish that excited states play a non negligible role in charge exchange processes for alkali ions on transition metal dichalcogenide surfaces and that they should be incorporated for quantitative theoretical descriptions. The findings also highlight the need for more sophisticated theoretical treatments incorporating dynamical correlation effects through coupled multi channel calculations.

Future work should concentrate on developing fully dynamical multi channel treatments that respect correlation effects throughout the trajectories, as well as extending analyses to other scattering geometries and projectile species.

\begin{acknowledgments}
This work was supported by CONICET through PIP-2021-101517 grant and U.N.L. through CAI+D grant No 85520240100061LI. We also wish to thank Edith C. Goldberg for her deep and valuable insights.
\end{acknowledgments}

\section*{AUTHORS DECLARATIONS}
\subsection*{Conflict of Interests}
The authors have no conflicts to disclose.
\subsection*{Data Availability Statement}
The data that support the findings of this study are available from the corresponding author upon reasonable request.

\bibliography{apssamp}

\end{document}